# Excitation wavelength-dependent ultrafast THz emission from surface and bulk of three-dimensional topological insulators

Anand Nivedan and Sunil Kumar*

*Femtosecond Spectroscopy and Nonlinear Photonics Laboratory,*
*Department of Physics, Indian Institute of Technology Delhi, Hauz Khas, New Delhi - 110016*
Email: *kumarsunil@physics.iitd.ac.in

## ABSTRACT

Three-dimensional topological insulators possess various interesting properties that are promising for various modern applications, including in the recently emerging fields of ultrafast THz photonics and spintronics. Their gapless spin-momentum-locked topological surface states with the presence of chiral spin structure are relevant for the development of light helicity-sensitive THz emitters and detectors. In this paper, we report femtosecond excitation pulse wavelength and helicity-dependent response of the three-dimensional topological insulators for an enhanced broadband THz pulse emission. Specifically, the excitation wavelength has been varied in a large range from near UV to near IR and two model systems of $Bi_2Te_3$ and $Bi_2Se_3$ single crystals have been studied. It was observed that the photoexcitation at shorter wavelengths enhances the THz emission from both the surface and bulk states. Primarily, circular photogalvanic effect is responsible for the surface contribution, while the prominent contribution in the bulk is from photon-drag effect. The wavelength-dependence of the THz emission is explained by accounting for the free carrier absorption of THz radiation within the material accurately.

# 1. INTRODUCTION

Three-dimensional topological insulators (3D-TIs) are a new class of quantum material systems which have become popular amongst researchers due to their unique electronic structure and hence provide a platform for discovering new science relevant for various modern applications. Bismuth telluride ($Bi_2Te_3$) and bismuth selenide ($Bi_2Se_3$), among others, are widely used and studied extensively using various experimental techniques to clearly distinguish the Dirac surface states in the overall electronic structure[1-3]. Various practical aspects of these materials for applications in thermoelectrics, optoelectronics and photovoltaics have been widely explored recently[4, 5]. Ultrafast techniques like time- and angle-resolved photoemission spectroscopy[6], optical pump-probe spectroscopy[7-9], THz emission and time domain spectroscopy[10-12] etc., help investigate the equilibrium and nonequilibrium textures of the charge as well as spin of the carriers and their excitations. Ultrafast optical excitation can induce nonlinear optical phenomena such as second harmonic generation (SHG)[13, 14], high harmonic generation (HHG)[15], nonlinear Kerr effect[16], photogalvanic effect (PGE)[17, 18], etc. on the surface, which can be very useful, especially in studying the characteristics of the Dirac Fermions in 3D-TIs. Strong spin-orbit coupling induced band inversion and time-reversal symmetry (TRS) are behind the origin of Dirac Fermionic states in TIs. While the TRS is protected on the surface of a 3D-TI and the fact that spin is locked with the direction of the electronic motion on the surface, back scattering of electrons on the surface is prohibited[19]. Therefore, 3D-TIs have become the desirable systems for low-power and high-speed spintronic and quantum electronic devices, even in the areas of quantum computing[20] and quantum communication[21].

Owing to the strong spin-orbit coupling property[22, 23], high spin-to-charge conversion efficiency in 3D-TIs is also being employed in spintronic applications at THz frequencies[24, 25]. Conversion of ultrafast spin current to ultrafast charge current via the inverse Edelstein effect (IEE) on the surface/interface and inverse spin Hall effect (ISHE) in the bulk[26, 27] are the underlying processes for ultrafast THz emission from heterostructures of 3D-TIs with highly spin polarizable material layers. It has been shown in the recent experimental studies that TI-based spintronic THz emitters can be as powerful as some of the most efficient ferromagnetic/nonmagnetic metallic heterostructures-based THz sources[11, 28, 29]. But the distinct feature of a 3D-TI-based THz emitter is in its ability to generate chiral THz waves from the conducting Dirac states present on its surface[30]. Since the Dirac states are localized only up to the thickness of ~2 nm at the surface[31], overall THz emission is dominated by the bulk contribution. Experimental studies have shown that various physical mechanisms are involved in the generation of THz radiation from 3D-TIs. Circular photogalvanic effect (CPGE) and linear photogalvanic effect (LPGE) are responsible for the generation of THz waves from the conducting Dirac states[30, 32], whereas the bulk contribution mainly comes from the photon-drag effect (PDE), photo-Dember effect, band-bending effect and photo-thermoelectric effect[12, 33-36]. Because these materials have such complicated THz emission pathways, various experimental techniques such as, thickness-dependent, carrier concentration-dependent and excitation helicity-dependent experiments are performed to separate the contribution of the surface from that of the bulk[37-39].

Since the surface states of 3D-TIs exhibit helical spin texture[40] with the spin of electrons locked to its momentum, an asymmetric excitation of carriers by circularly polarized light is necessary to achieve spin-polarized current[17, 41-43]. In a recent study, differential selectivity of spin-up and spin-down electrons by circularly polarized light has been demonstrated for the generation of chiral THz radiation from the surface of 3D-TIs[30]. Much more experiments are required to exploit this further, knowing the fact that the overall THz emission efficiency of these materials is limited by the free carrier absorption (FCA) of THz waves inside the samples[44]. The as-grown $Bi_2Te_3$ family of 3D-TIs are weakly metallic at room temperature due to the intrinsic doping introduced during the growth and hence they would provide a weaker THz emission relative to otherwise ideally insulating in the bulk counterparts. Enhancing CPGE and LPGE in 3D-TIs is important for stronger surface contribution to THz radiation. It has been seen that tuning of Fermi level with respect to Dirac point enhances photogalvanic current in these materials[45]. Luo et al. reported a significant improvement in THz emission from Cu-doped $Bi_2Se_3$ by suppressing high degree of intrinsic carrier concentration[44]. However, tuning of Fermi level via counter doping or electrical gating results in the decreased mobility of the carriers, especially on the surface states[46]. And it has been seen that the mobility of carriers greatly affects the strength of emitted THz radiation[28]. In general, high electron mobilities in narrow band-gap semiconductors are thought to be one of the reasons for their strong THz emission efficiency[47, 48]. Therefore, it is desirable to search for alternative non-invasive techniques to improve THz emission from surface states of 3D-TIs.

Here in this work, we investigate excitation wavelength and helicity-dependent THz emission from 3D-TIs, particularly the $Bi_2Te_3$ and $Bi_2Se_3$ single crystals. THz emission time-domain spectroscopy experiments have been performed following ultrafast excitation by amplified femtosecond laser pulses with varying polarization and central wavelength in a large range from near UV to NIR. We show that the THz emission efficiency increases with the decreasing excitation wavelength. Further, distinguishable contributions to THz emission from the surface and the bulk states are extracted and analysed for all the experimental conditions. The surface contribution to THz emission is seen to be increasing towards the shorter wavelengths more rapidly than the bulk contribution. A new scheme is proposed in the paper to interpret the observations accurately by appropriately accounting for the loss of THz radiation due to free carrier absorption within the optically excited region in the sample under study.

## 2. MATERIALS AND METHODS

*THz time-domain spectroscopy:* **Figure 1a** shows schematically the experimental setup used for THz emission spectroscopy in reflection geometry. The sample was irradiated with pulses of time-duration of ~50 fs and centred at 800 nm from a Ti:sapphire based regenerative amplifier (Astrella, Coherent Inc. USA) operating at 1 kHz pulse repetition rate. The laser output was divided into two parts using a beam splitter (BS); the stronger one was routed through an optical parameter amplifier (OPA, Opera Solo, Light Conversion, USA) and used as the pump in the experiment, while the second part was routed through a time-delay stage and used for detection of THz pulse emitted by the sample. Excitation pulses with varying central wavelength from near UV to near infrared (NIR) were produced inside the OPA. Nearly collimated excitation (pump) beam of diameter ~1.6 mm and fluence of ~150 µJ/cm$^2$ was used for all the results reported in this paper. The emitted THz pulses were collected by a 15 cm focal length off-axis parabolic mirror (PM1), which are then focused onto a nonlinear crystal with the help of another similar parabolic mirror (PM2). The optical beam for detection is routed through a small hole in PM2 and is made collinear with the THz beam onto the nonlinear crystal. THz pulse detection is done by electro-optic sampling scheme which combines the use of a quarter wave plate, a Wollaston prism and a balanced photodiode in conjunction with a lock-in amplifier. More details about the detection can be found elsewhere[29]. The THz emission experiments in the current study have been performed for different excitation helicity (α) varied by a quarter-wave plate and azimuthal angle (ϕ) variable by rotating the sample as shown in **Fig. 1b** keeping the incident angle (θ) constant at ~45°.

*Material*: Bismuth telluride family of 3D topological insulators crystallize in a rhombohedral (trigonal) unit cell structure recognized by $R\bar{3}m$ space symmetry group[49, 50]. A schematic of the atomic structure of $Bi_2(Te, Se)_3$ is shown in **Fig. 1c**. Atoms in these compounds are covalently bonded together to form a quintuple layer and each quintuple layer binds together along c-axis direction by weak Van der Waal forces to form a hexagonal lattice. Since these are layered materials, these can be mechanically exfoliated to get thin flakes of very high surface quality. The figure shows the basal plane of a 3D topological insulator. It possesses inversion symmetry, threefold rotation around the c-axis, and a mirror plane symmetry in the bulk. Therefore, bulk of these compounds is centrosymmetric in nature[51, 52]. However, due to crystal termination at the boundary, space inversion symmetry is broken. As a result, THz emission due to second-order nonlinear processes like optical rectification (OR), CPGE and LPGE are possible on the surface of 3D TIs[53, 54]. This crystalline nature of 3D-TIs allows us to explore the properties of the Dirac Fermions of even highly doped samples whose weak metal-like transport properties are dominated by doping in the bulk states.

**Figure 1d** shows typical THz waveforms generated from the samples following ultrafast optical excitation at 800 nm using either LP ($\alpha = 0°$) or LCP ($\alpha = -45°$) or RCP ($\alpha = +45°$) light. The difference in the THz signal magnitude is apparent from this representation, which along with complete details with respect to the varying excitation wavelength and azimuthal angle, have been discussed below in the Results and discussion section. Schematic of band structure of a 3D-TI, as shown in **Fig. 1e,** indicates the differences in the optical excitation of carriers on the surface by different helicity light. When a 3D-TI is illuminated with a linearly polarized light, a net spin current is produced on the surface, but the charge current remains zero[55]. Thus, an asymmetric excitation of carriers is necessary to achieve spin-polarized current to contribute a net charge current uniquely on the surface of 3D-TIs. Circularly polarized light having specific spin-angular momentum selectively excites electrons of a particular spin on the surface of a 3D-TI, as indicated in Fig. 1e[41, 42]. This asymmetrical excitation of electrons leads to the generation of a net charge current on the surface. Thus, for a time-dependent charge current, which is the case with ultrafast photoexcitation, pulses of THz radiation are emitted.

## 3. RESULTS AND DISCUSSION

*3.1. Pump-helicity and wavelength-dependent THz emission*

**Figure 2** summarizes the main results of pump-helicity and excitation wavelength-dependent THz emission in reflection geometry from freshly cleaved samples of $Bi_2Te_3$ and $Bi_2Se_3$. The helicity-dependent THz experiments are helpful in separating individual contributions due to nonlinear CPGE and PDE processes, and the same is presented in Figs. 2a-c and 2d-f for $Bi_2Te_3$ and $Bi_2Se_3$, respectively. Another nonlinear effect, i.e., optical rectification (OR), contributes relatively negligibly to the THz emission from the surface of a 3D-TI[30, 31]. We have verified this fact through azimuthal angle-dependent experiments. The corresponding results using linearly polarized light at 800 nm are presented in Fig. 2g, and the analysis is given in the following paragraphs. The three-fold rotational symmetry in the crystal of $Bi_2Te_3$ and $Bi_2Se_3$, as indicated in Fig. 1c, is responsible for the results in Fig. 2g[37]. For achieving maximum nonlinear contribution to THz emission, the crystal azimuthal angle was set for maximum overall THz signal.

The actual THz waveforms, as recorded in the experiments at the three different polarizations and at fixed azimuthal angle corresponding to maximum signal, are shown in **Figs. 2a and 2d** for $Bi_2Te_3$ and $Bi_2Se_3$, respectively, where the results at different wavelengths are shifted horizontally for clarity. Three observations are apparent: (i) the magnitude of the THz signal is the largest for LP at all wavelengths, (ii) the magnitude of the THz signal increases at shorter wavelengths for all polarizations, and (iii) the signal from $Bi_2Te_3$ is nearly five-times stronger than that from $Bi_2Se_3$ at the shortest experimental wavelength. The peak-to-peak value of the THz electric field ($E_{pp}$) as a function of the pump wavelength has been plotted in **Fig. 2b** for $Bi_2Te_3$ and **Fig. 2e** for $Bi_2Se_3$ for all the three helicities of the pump. At a given value of α, the THz signal has multiple contributions, which can be analysed through the following phenomenological equation[17, 32, 38, 41, 42],

$$E_{pp}(\alpha) = C\sin(2\alpha) + L\sin(4\alpha) + P\cos(4\alpha) + O \qquad (1)$$

Here, on the right-hand side, the first term having coefficient $C$ is contributed by spin-selective CPGE process, the second and the third terms are contributions from spin-independent processes, i.e., LPGE and PDE, respectively. The fourth term represents the remaining THz contribution coming from all other effects that are helicity-independent in nature. CPGE and LPGE are second-order nonlinear optical effects which occur on the topological surface, whereas the remaining from the above-mentioned effects, including PDE, mainly occur in the bulk[12, 33-35]. From our experimental results for LCP and RCP excitations, the contribution of CPGE can be extracted using the relation, $E_{pp}(CPGE) = (1/2)\{E_{pp}(\alpha = -45°) - E_{pp}(\alpha = 45°)\}$ (deduced from equation 1 above)[56]. Spin-independent process, LPGE does not contribute to THz emission for α = 0°, 45° and -45° *clf*, the second term in Eq. (1). Similarly, the contribution of PDE in the THz emission can be extracted from Eq. (1) as $E_{pp}(PDE) = (1/4)\{2E_{pp}(\alpha = 0°) - E_{pp}(\alpha = -45°) - E_{pp}(\alpha = 45°)\}$. The detailed derivation for these two mechanisms has been given in **Section S3, Supporting Information.** Both the CPGE and PDE contributions in the THz signals from $Bi_2Te_3$ have been presented in **Fig. 2c** as a function of the excitation wavelength. Quite consistently, same observations are found for $Bi_2Se_3$ as shown in **Fig. 2f**. The magnitude of the THz signal from $Bi_2Se_3$ at longer wavelengths, beyond ~900 nm was significantly weak and within the experimental noise, hence the same could not be analysed in Fig. 2f.

As mentioned earlier, azimuthal angle-dependent THz emission experiments help to extract the optical rectification contribution from within the ensemble of all the nonlinear contributions. Here, we show that there is an insignificant contribution from optical rectification at the surface of $Bi_2Te_3$ and $Bi_2Se_3$. This was found to be true for all the polarizations and different excitation wavelengths. The ϕ-dependence of the THz signal recorded for linear polarization of 800 nm excitation is presented in **Fig. 2g**. Clearly, 3-fold rotational symmetry is observed as expected. The oscillatory component in the response usually has origin from the nonlinear effects [30, 37], including the optical rectification, while the constant offset of the response can arise cumulatively from the photo-Dember effect, band-bending effect, photo-thermoelectric effect, etc., in the bulk, all represented together by the term 'O' in Eq. 1. Under a fixed helicity excitation, the overall contribution of linear and nonlinear effects can be given by

$$E_{pp}^{total}(\phi) = E_{pp}^{linear} + E_{pp}^{nonlinear}(\phi) \equiv A + B\cos(3\phi) \quad (2)$$

where, coefficients A and B represent the two contributions, respectively. The extracted value of the oscillatory nonlinear term, B (at ϕ = 0) matches well with the PDE contribution to the THz emission as extracted from the helicity-dependent analysis presented earlier in the paper, individually for both the crystals. Additionally, we observe that the ratio between the linear and the nonlinear components for both the $Bi_2Se_3$ (A/B ~ 4) and $Bi_2Te_3$ (A/B ~ 3.5) are nearly same.

Having the above analysis done from the helicity and azimuthal angle dependent THz emission measurements, we are now able to quantify the THz emission, contributed by the surface states and the bulk states, separately. As is clear now, the surface contribution is mainly from CPGE, which after subtracting from the overall THz emission, would provide the contribution from the bulk. These two have been calculated for RCP excitation light at all wavelengths and plotted as shown in **Fig. 2h** for both $Bi_2Te_3$ and $Bi_2Se_3$. Notably, both the surface and bulk components in the THz emission are increasing with the decreasing wavelength. From **Fig. 2i**, where the ratio between the two parts, $E_{pp}^{surface}/E_{pp}^{bulk}$, is plotted, we conclude that the enhancement in the surface component relative to the bulk is stronger for the shorter wavelengths. Higher THz signal from surface states at shorter wavelengths or higher photon energies is consistent with the fact that high energy photons which are more surface sensitive are used to determine only the surface band structure of 3D-TIs using ARPES[6].

*3.2. Effect of FCA on THz emission*

In this section, we discuss the role of intrinsic free carriers on the THz emission efficiency of $Bi_2Te_3$ and $Bi_2Se_3$ crystals. Since the two crystals used in our experiments are metallic in nature (**Section S2, Supporting Information**), THz waves are absorbed by the large number of free carriers present within the material due to intrinsic doping[57, 58]. For incorporating the effect of FCA, two possible schemes can be thought of. In the first that has been used in the literature[44], one calculates the attenuation of THz radiation by FCA using the standard formula where the THz source is assumed to be placed below the surface at a depth equal to the optical penetration depth. This scheme fails to explain our observations at longer excitation wavelengths. In fact, the exponential decay of the excitation intensity below the surface is also ignored in this scheme. In the second or the modified scheme, as we propose here, one considers a continuous series of THz source points below the surface and mostly within the optical penetration depth in the photoexcited volume that contribute to the overall THz emission. The latter one is more appropriate as it eliminates all the inconsistencies and fits well with the observations. Both are now discussed in more detail in the following paragraphs.

In **Fig. 3a**, we show, schematically, the process of illumination of the sample surface by a femtosecond NIR pulse and the subsequent generation of THz waves from a point located at the optical penetration depth ($\delta$) inside. This is the formulation of the scheme-I discussed in the above paragraph. For a thick crystal (thickness ≫ optical penetration depth), like one in the present study, THz emission is measured in the epi-direction and therefore, the THz attenuation by FCA within $\delta$ must be considered. Assuming $E_{pp}^o$ as the magnitude of the THz signal generated at the source point at $\delta$ and having the THz attenuation by FCA included, the measured THz signal outside the sample surface, $E_{pp}$ can be given by the relation[44],

$$E_{pp} = E_{pp}^o \, e^{-\frac{\sigma}{2}N\delta} \qquad (3)$$

where $\sigma$ is the THz absorption cross-section and $N$ is the uniform free carrier density. See **the Supporting Information** for the values. Using the above relation, we have calculated $E_{pp}^o$ which represents the magnitude of the THz signal on the detector outside if FCA was not considered. The $E_{pp}$ is known from the experiments as discussed earlier and shown in Figs. 2b and 2e for all the experimental wavelengths and helicity of the excitation light. Thus, calculated values of $E_{pp}^o$ for the experimental data shown in Figs. 2b and 2e for the linearly polarized light excitation are presented in **Figs. 3b and 3c**. We observe that $E_{pp}^o$, i.e., the magnitude of the THz signal generated at the source point at $\delta$ inside the crystal, increases with the increasing wavelength for both the samples and becomes unrealistically high at the longer wavelengths. See the THz electric field calculation in **Section S7 of the Supporting Information**. This observation necessitates a modification in the scheme that has been used to analyse the data.

Now we come to the modified scheme as represented by the schematic shown in **Fig. 3d** for the optical excitation of the sample and THz emission from it. In the model, we consider several photoexcited points within the excitation volume along the pump beam propagation as potential sources of THz radiation. The magnitude of the THz radiation produced by ith representative point within a differential thickness, $\Delta d = d_i - d_{i+1}$, as shown in Fig. 3d, depends on the exponentially decayed excitation intensity in that region and is given as

$$\Delta E_{pp}^{d_i} = E_{pp}^{d_i} - E_{pp}^{d_{i+1}} \equiv E_{pp}^o \, e^{-d_i/\delta} - E_{pp}^o \, e^{-d_{i+1}/\delta} \qquad (4)$$

The field $E_{pp}^{d_i}$ accounts for the THz emission from all the points below $d_i$ within the photoactivated thickness in the sample. In the above expression, $E_{pp}^o$ is the total THz radiation as it is produced. Since the region in which the THz sources are distributed is extremely narrow as compared to the wavelength of the generated THz radiation, hence radiation from all the differential thicknesses add up coherently at the detector outside. Total THz radiation as generated by all the source points is obtained by summing over all of them and is given as

$$E_{pp}^o = \sum_{i=1}^{\infty} \Delta E_{pp}^{d_i} \equiv \sum_{i=1}^{\infty} \left( E_{pp}^o \, e^{-d_i/\delta} - E_{pp}^o \, e^{-d_{i+1}/\delta} \right) \qquad (5)$$

In principle, the source points are distributed all along the extended depth, even below the penetration depth in the sample. However, most of the THz radiation is contributed by those within the optical penetration depth. By including the FCA, the magnitude of the emitted THz radiation to be measured at a detector outside the sample can be given by the relation,

$$E_{pp} = E_{pp}^o \left\{ \sum_{i=1}^{\infty} \left( e^{-d_i/\delta} - e^{-d_{i+1}/\delta} \right) e^{-\frac{\sigma}{2} N d_i} \right\} \qquad (6)$$

In the above relation, the value of $E_{PP}$ is inserted from the experimentally obtained results and hence, $E_{pp}^o$ is estimated at all wavelengths. The corresponding results for $E_{PP}$ and $E_{pp}^o$ are shown in **Fig. 3e and 3f** for Bi$_2$Te$_3$ and Bi$_2$Se$_3$, respectively. Clearly, irrespective of the excitation wavelength, the generated THz signal is of the same magnitude. Having smaller value of $E_{pp}^o$ from Bi$_2$Se$_3$ as compared to that from Bi$_2$Te$_3$ is possibly due to the difference in the carrier mobility in these two systems[28].

From magneto-transport measurements, we have determined various physical properties of these two systems as summarized in **Table 1** of **Supporting Information**. It is noteworthy to mention a few notable points here from the comparison in the behaviour of the two crystals with regards to the THz emission from them and the magneto-transport experiments. While Bi$_2$Te$_3$ has carrier mobility nearly 1.5 times larger than that in Bi$_2$Se$_3$ (see **Table 1** in the **Supporting Information**), the magnitude of the THz radiation, $E_{pp}^o$ generated from Bi$_2$Te$_3$ is found to be nearly 2.5 times stronger than that from Bi$_2$Se$_3$. This observation will be helpful in future to formulate a relationship between the THz generation efficiency and the material parameters and hence device suitable material structures for customized THz emitters based on 3D-TIs.

## 4. CONCLUSION

To summarize, an enhanced THz emission from the surface and bulk states of 3D-TIs following ultrafast optical excitation at shorter wavelengths is reported here. Systematic experiments with different helicity of the excitation light and varying azimuthal angles on single crystals of $Bi_2Te_3$ and $Bi_2Se_3$ help to unravel distinct contributions from both the surface and the bulk states. We find that the bulk has a significant contribution from photon-drag effect when excited by either circularly or linearly polarised light, whereas the surface contribution primarily originates due to CPGE in the presence of circularly polarised light only. As compared to the bulk, the surface contribution to THz emission gets enhanced more rapidly towards shorter wavelengths. To correctly account for the excitation wavelength-dependent variation in the magnitude of the generated THz radiation vis-à-vis the one after incorporating free-carrier absorption, a new model has been proposed in the paper, which in general, can be applicable to any other material system. Apart from the wavelength dependence, it has been discovered that the intrinsic THz generation efficiency of $Bi_2Te_3$ having higher carrier mobility, is superior to that of $Bi_2Se_3$. Stronger chiral response of the 3D-TIs in terms of spin-selective excitation at shorter wavelengths, as observed in our study, can be highly relevant for developing chirality-sensitive efficient THz emitters and detectors in future.

**Supporting Information**

Supporting Information is available.

**Acknowledgements**

SK acknowledges the Science and Engineering Research Board (SERB), Department of Science and Technology, Government of India, for financial support through project no. CRG/2020/000892. Joint Advanced Technology Center, IIT Delhi is also acknowledged for support through EMDTERA#5 project. Central Research Facility, IIT Delhi is acknowledged for PPMS facility. AN acknowledges DST for INSPIRE Fellowship.

**Conflict of Interest**

The authors declare no conflict of interest.

## FIGURE CAPTIONS

**Fig. 1** THz emission measurements on ultrafast photoexcited single crystals of three-dimensional Bi$_2$Te$_3$ and Bi$_2$Se$_3$ topological insulators. Schematic of a) the experimental setup arranged for measuring the emitted THz signal in the reflection geometry and by varying b) excitation light helicity α as defined by the QWP angle, the incident angle, θ from the crystal surface normal, i.e., C-axis, and the azimuthal angle, ϕ. c) Layered nature of the crystals represented in a quintuple layer and the basal plane to indicate the high-symmetry $\sigma_v$-plane and C$_3$ rotational axis (C-axis). d) Experimentally measured THz waveforms under excitation by LP (α = 0°), LCP (α = -45°) and RCP (α = +45°) light. e) Band structure of a 3D-TI depicting the excitation of carriers from the surface states by linearly and circularly polarized light. QWP: quarter wave plate; LP: linearly polarized; LCP: left circularly polarized; RCP: right circularly polarized; BS: beam splitter; PM: parabolic mirror, ZnTe: zinc telluride; WP: Wollaston prism; BPD: balanced photodiode.

**Fig. 2** Wavelength-dependence of magnitude of THz radiation emitted by Bi$_2$Te$_3$ and Bi$_2$Se$_3$ at varying helicity of the excitation light corresponding to α = 0, +45 and -45. a-c) Experimental results for Bi$_2$Te$_3$. a) Raw-data from the time-domain THz emission measurements. Data at different excitation wavelengths shifted horizontally for clarity. b) Estimated values of the peak-to-peak electric field, $E_{pp}$ of the THz pulse, and c) the contributions from the PDE and CPGE (see text for details). d-f) Experimental results for the Bi$_2$Se$_3$ crystal consistent with the results for Bi$_2$Te$_3$. g) $E_{pp}$ at varying azimuthal angle to determine the contribution from surface optical rectification. Continuous curves are fitting of the data using Eq. (1). h) Absolute surface and bulk contributions to the THz generation, and f) ratio of surface to bulk contributions indicating rapid increase in the THz generation from the surface carriers at shorter wavelengths.

**Fig. 3** Wavelength-dependent THz emission and the effect of FCA in 3D-TIs excited by linearly polarized femtosecond laser pulses. a) Scheme-I considering THz generation from a point that is a penetration depth, $d_\lambda$ underneath the surface. Amplitude of the THz pulse generated inside ($E_{pp}^o$) and that measured outside ($E_{pp}$) the crystal at varying excitation wavelengths in Scheme-I for b) Bi$_2$Te$_3$ and c) Bi$_2$Se$_3$. d) Scheme-II considering THz generation at various points within the penetration depth inside the crystal. e,f) Amplitude of the resultant THz pulse inside ($E_{pp}^o$)

obtained by integrating the emission at all points within the penetration depth and that measured outside ($E_{pp}$) the crystal at varying excitation wavelengths in Scheme-II. The horizontal lines in e) and f) represent constant mean values of $E_{pp}^o$ for the two crystals.

**FIGURE 1**

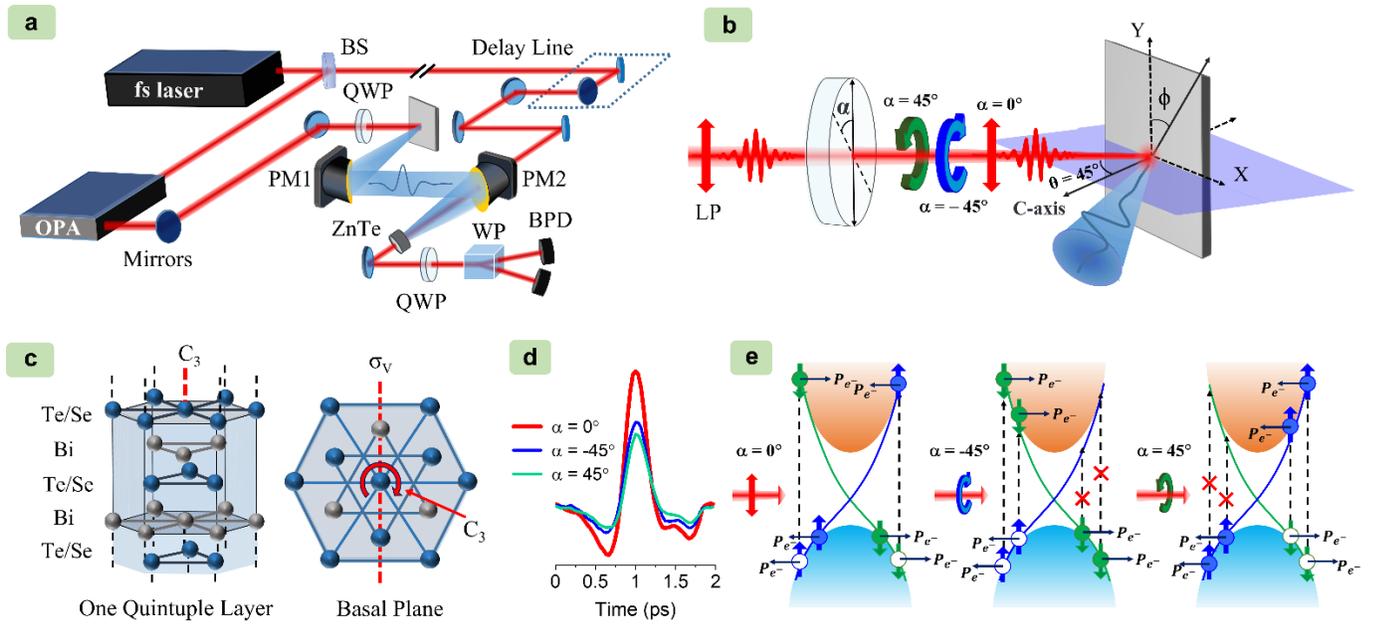

**FIGURE 2**

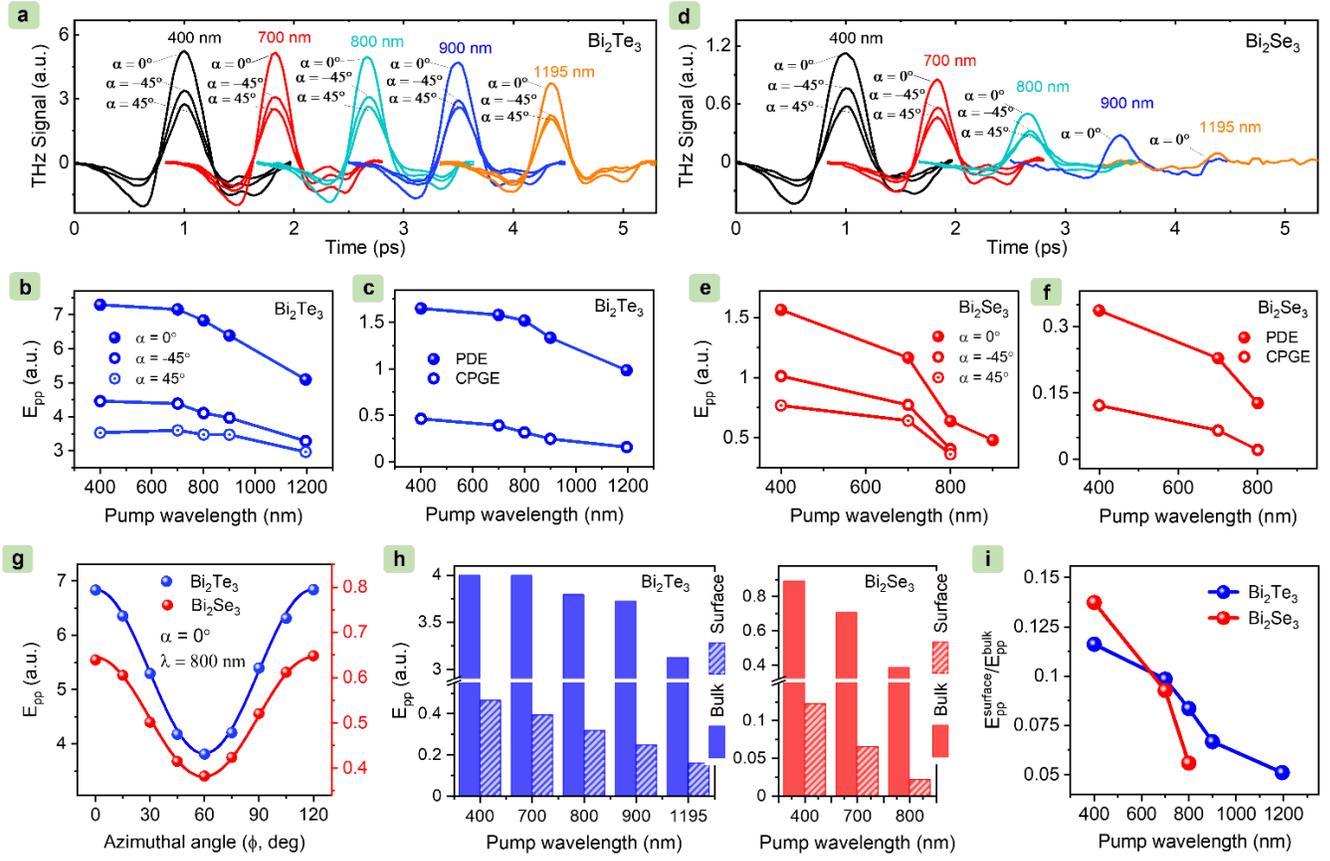

**FIGURE 3**

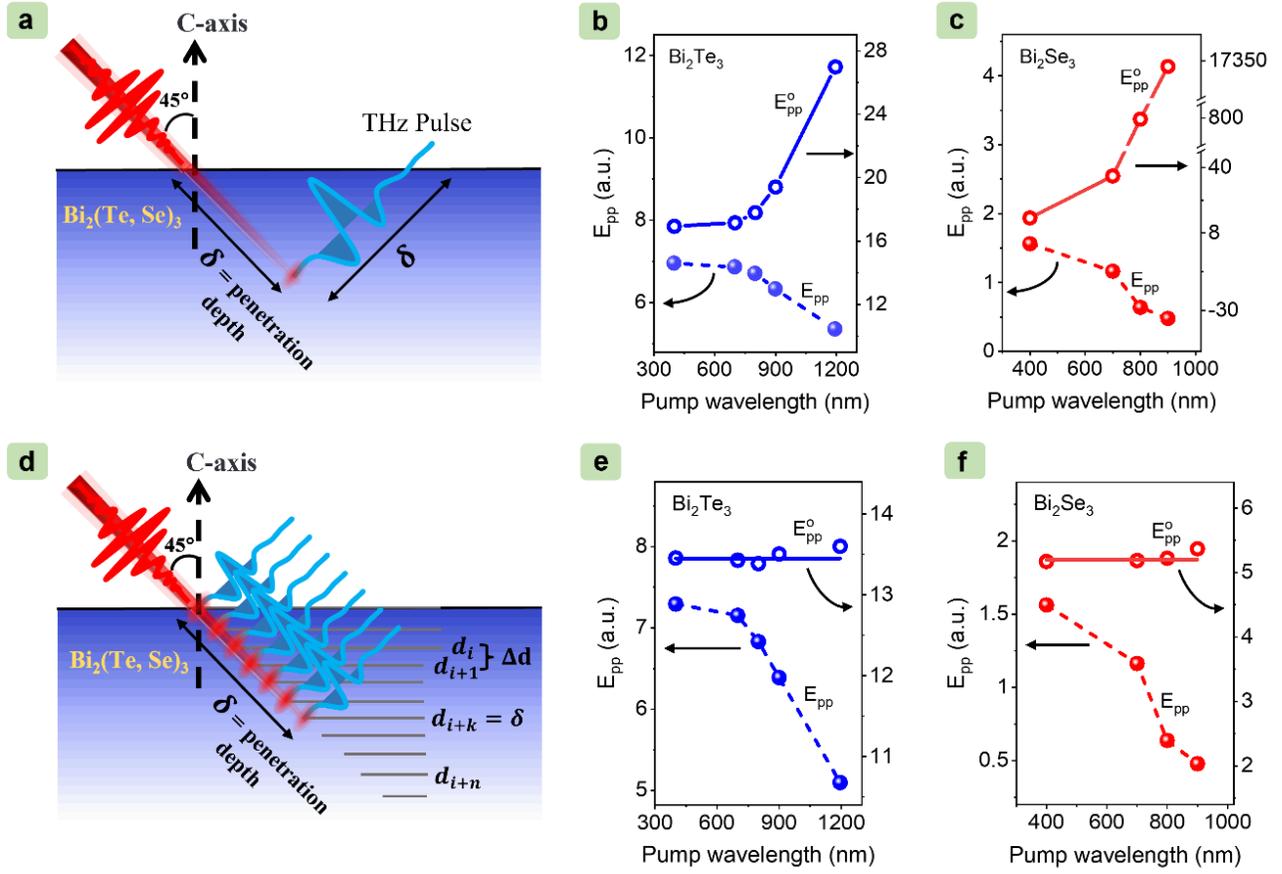

# Excitation wavelength-dependent ultrafast THz emission from surface and bulk of three-dimensional topological insulators

Anand Nivedan and Sunil Kumar*

*Femtosecond Spectroscopy and Nonlinear Photonics Laboratory,*
*Department of Physics, Indian Institute of Technology Delhi, Hauz Khas, New Delhi - 110016*
*Email: \*kumarsunil@physics.iitd.ac.in*

**S1. Raman and XRD spectra of $Bi_2Te_3$ and $Bi_2Se_3$**

Samples used in our experiments are of very high quality and single crystalline in nature. We used Raman scattering and X-ray diffraction (XRD) measurements to characterize these samples (**Fig. S1a and S1b**). All the experiments were performed on the thin flakes of $Bi_2Te_3$ and $Bi_2Se_3$, freshly cleaved from the bulk crystals. The presence of three Raman active modes, one in-plane ($E_g^2$) and two out-of-plane ($A_{1g}^1$ and $A_{1g}^2$) at ∼102 cm$^{-1}$, ∼62 cm$^{-1}$ and ∼134 cm$^{-1}$, respectively, in case of $Bi_2Te_3$, and at ∼131 cm$^{-1}$, at ∼72 cm$^{-1}$ and ∼174 cm$^{-1}$, respectively, in case of $Bi_2Se_3$ is consistent with the literature[1]. The **Fig. S1b** shows narrow XRD peaks of the samples with FWHM of ∼ 0.3° with the presence of only (0, 0, 3$n$) diffraction peaks in the XRD spectrum. These two features indicates that the films are crystalline and highly oriented along the c-axis.

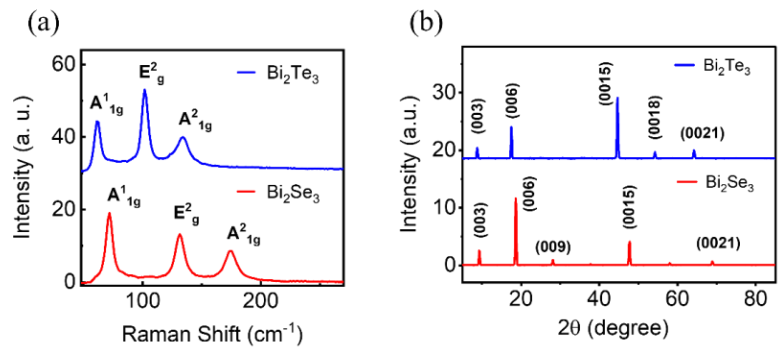

**Figure S1.** a) Raman spectra of the single crystalline $Bi_2Te_3$ and $Bi_2Se_3$ with all the three vibrations modes. b) XRD spectra of the single crystalline $Bi_2Te_3$ and $Bi_2Se_3$ showing the presence (0, 0, 3$n$) diffraction peaks.

## S2. Transport measurements

Temperature-dependent resistivity and Hall measurements are done on freshly cleaved $Bi_2Te_3$ and $Bi_2Se_3$ samples (**Fig. S2**) in cryogen free measurement system (CFMS) from Cryogenic Limited. **Fig. S2a** shows the temperature-dependent resistivity measurement (B=0) for $Bi_2Te_3$ and $Bi_2Se_3$ in the temperature range of 2-300 K. The resistivity of both the crystals increases with temperature, indicating metal-like behaviour in the given temperature range. **Fig. S2b** shows the Hall effect data for $Bi_2Te_3$ and $Bi_2Se_3$ in the magnetic field range of -4 to 4 T carried out at the room temperature. The negative slope of $Bi_2Se_3$ indicates that electrons are the majority charge carriers in the sample, in contrast to the slope of $Bi_2Te_3$, which is positive and indicates that the sample exhibits p-type behaviour. From the data, we determine the room temperature carrier concentration of $Bi_2Te_3$ and $Bi_2Se_3$ to be $3.43 \times 10^{19}$ and $14.09 \times 10^{19}$, respectively. Data analysis from the transport measurements indicates that the samples under study are primarily metallic in nature. This is consistent with experimentally measured transport properties of as-grown single crystals of $Bi_2Te_3$ and $Bi_2Se_3$ reported in the literature[2-4]. In general, $Bi_2Te_3$ family of 3D-TIs synthesized via all major techniques suffer from the same problem of extremely high value of intrinsic carrier concentration due to vacancies and antisite defects[5-7].

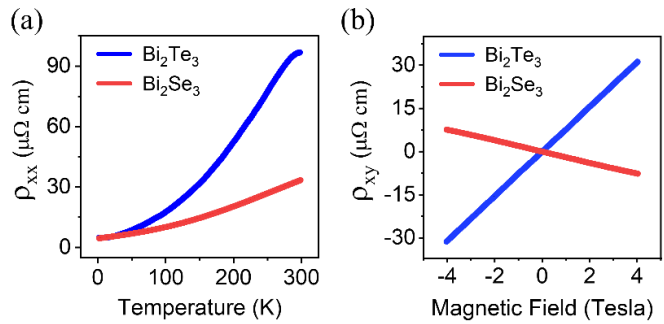

**Figure S2.** Transport measurement on $Bi_2Te_3$ and $Bi_2Se_3$. a) Temperature-dependent resistivity showing metal-like nature of the samples. b) Experimentally measured Hall resistivity indicating naturally grown p-doped $Bi_2Te_3$ and n-doped $Bi_2Se_3$

**Table 1.** Physical properties of $Bi_2Te_3$ and $Bi_2Se_3$ extracted from resistivity and the Hall-effect measurements at room temperature.

| Material | $t$ (μm) | $R_H$ (μΩ cm/Tesla) | $N$ ($cm^{-3}$) | $\rho_{xx}$ (μΩ cm) | $\mu$ ($m^2$/V s) | $\tau$ (ps) |
|---|---|---|---|---|---|---|
| $Bi_2Te_3$ | 130 | 18.23 | $3.43 \times 10^{19}$ | 96.83 | 0.1883 | 0.15 |
| $Bi_2Se_3$ | 180 | 4.46 | $14.09 \times 10^{19}$ | 33.40 | 0.1335 | 0.11 |

## S3. Phenomenological model of photocurrent generation in 3D topological insulators

Transient photocurrent can be induced in 3D-TIs by a variety of mechanisms. To understand the underlying principles behind the THz emission from 3D-TIs it is extremely important to distinguish these processes which induce current on the surface and the in the bulk of 3D-TIs. Polarization-dependent excitation is one of the tools to realise such phenomena. The characteristic polarization-dependent behaviour of persistent current[8-10] in transport measurements or THz emission due to transient current in THz spectroscopy measurements[11] in 3D-TIs has been phenomenologically described by the relation,

$$E_{pp}(\alpha) = C\sin(2\alpha) + L\sin(4\alpha) + P\cos(4\alpha) + O \quad (S3.1)$$

Here, the first term having coefficient $C$ represents the contribution of spin-selective CPGE process, the second and the third terms are contributions from spin-independent processes, i.e., LPGE and PDE, respectively. The remaining THz contribution from all other helicity-independent effects is represented by the fourth term $O$. For the excitation of the sample with the linearly polarized laser beam, equation 1 reduces to $E_{pp}(\alpha = 0°) = C\sin(0°) + L\sin(0°) + P\cos(0°) + O$, i.e., $E_{pp}(LP) = P + O$. For LCP excitation, equation 1 reduces to $E_{pp}(\alpha = -45°) = C\sin(-90°) + L\sin(-180°) + P\cos(-180°) + O$, i.e., $E_{pp}(LCP) = -C - P + O$. Similarly, For RCP excitation, equation 1 reduces to $E_{pp}(\alpha = 45°) = C\sin(90°) + L\sin(180°) + P\cos(180°) + O$, i.e., $E_{pp}(LCP) = C - P + O$. From these three relations, we note that the contribution of CPGE and PDE can be separated even if the experiments are performed just for three polarizations. The contribution of CPGE in THz emission can be extracted by subtracting $E_{pp}(RCP)$ from $E_{pp}(LCP)$. Similarly, the contribution of photon-drag effect can be extracted by the relation $E_{pp}(PDE) = (1/4)\{2E_{pp}(\alpha = 0°) - E_{pp}(\alpha = -45°) - E_{pp}(\alpha = 45°)\}$. Notably, the spin-independent process LPGE does not contribute to THz emission for α = 0°, 45° and -45° excitations. To account for the THz emission due to LPGE, experiments are needed to be performed for other polarizations also. Subsequently, the contribution of LPGE can be separated by extracting the value of fitting parameter L for the experimental results.

Circular photogalvanic effect (CPGE) is the main contribution of THz radiation from the surface of $Bi_2Te_3$ and $Bi_2Se_3$ in our experimental results. Phenomenological equation describing the transient current due to CPGE is given by[12-14],

$$j_i = \sum_j \gamma_{ij}^{(2)} (\vec{E} \times \vec{E}^*)_j \quad (S3.2)$$

where $\vec{E}$ is the electric field vector of the incident light, $\gamma_{ij}^{(2)}$ is second-rank pseudo tensor. Since the magnitude of current due to CPGE reverses its sign upon changing the helicity of the incident light polarization, the relation in S3.2 can be expressed as $j \propto L_{polz} E_o^2$ where $E_o$ represents the amplitude of the electric field of the pump excitation, and $L_{polz}$ denotes the degree of circular polarization of the electromagnetic waves. For RCP, $L_{polz}$ represents the highest degree of polarization with the value of +1, and for LCP the value of $L_{polz}$ becomes -1. Variation of the polarization of the pump beam is changed using a QWP that changes the polarization of light that follows $L_{polz} = \sin(2\alpha)$,[14] where α is defined as the angle between the plane of polarization of the input beam and the optical axis of the QWP.

This gives a relation between the transient circular photogalvanic current and the angle of QWP rotation as shown in equation S3.1.

The photon drag effect on the other hand is caused by the momentum transfer of photons to charge carriers. When light is absorbed by free carriers, momentum of the electromagnetic wave is transferred to the carriers along with its energy. As a result, excited carriers acquire translational motion causing formation of current in the materials. Phenomenologically, photon drag effect is given by[15],

$$j_i = \sum_{j,k,l} \chi^{(2)}_{ijkl} k'_j E_k(\omega) E^*_l(\omega) \tag{S3.3}$$

wher $\chi^{(2)}_{ijkl}$ is the fourth rank tensor and $k'$ is the momentum wave vector of the incident radiation. Microscopically, photons can excite electrons from initial state $\Psi_i(k_0)$ to the final state $\Psi_f(k_0 + k')$, where momentum $k'$ of photons is transferred to the excited electrons.

## S4. Wavelength-dependent penetration depth

In general, penetration of EM waves increases with the wavelength. For $Bi_2Te_3$ and $Bi_2Se_3$, penetration depth of light in the wavelength range of 400-1200 nm has been reported to be 23-47 nm and 17-170 nm, respectively[16].

Table 2. Penetration depth of excitation light in the crystals[16].

| Excitation wavelength (nm) | 400 | 700 | 800 | 900 | 1195 |
|---|---|---|---|---|---|
| Photon energy (eV) | 3.10 | 1.77 | 1.55 | 1.38 | 1.04 |
| Penetration depth for $Bi_2Te_3$ (nm) | 23.5 | 24.4 | 26.7 | 31 | 46.5 |
| Penetration depth for $Bi_2Se_3$ (nm) | 17 | 25.5 | 53 | 78 | 169 |

## S5. Calculation of absorption cross-section (σ)

We calculate absorption cross section using the relation $\sigma = \frac{\mu e}{(c n_r \varepsilon_o)(1+ \omega^2 \tau^2)}$, where µ is the mobility of the carriers, c is the speed of light in vacuum, $n_r$ is the refractive index of the material, $\varepsilon_o$ is the permittivity of free space, $\omega$ is the angular frequency of the THz wave, and τ is the relaxation time of the carriers. Mobility, $\mu = \frac{1}{\rho_{xx} N e}$, of $Bi_2Te_3$ and $Bi_2Se_3$ is evaluated from the resistivity and the Hall effect measurement data at room temperature, which has been summarised in **Table 1** (**Section S2**). The peak angular frequency obtained from the THz spectrum (**Section S5**) has the value of ~ 1 THz for both samples. We observe that the absorption cross section for the whole THz spectrum from 0.1 to 2.5 THz is almost invariable. With the effective mass of the electron, m ≈ 0.14mo[17, 18], relaxation time $\tau = \frac{\mu m}{e}$ of $Bi_2Te_3$ and $Bi_2Se_3$, is calculated to be 0.15 and 0.11 ps respectively. It has been reported that the refractive index of both these samples is ~3 in the frequency range of up to 2 THz[19]. The calculated value of σ for $Bi_2Te_3$ is $2.09 \times 10^{-14}$ and for $Bi_2Se_3$ is $1.91 \times 10^{-14}$ which is of the same order as obtained in the previous experimental result by Luo et al.[20].

## S6. THz waveform and spectrum

**Fig. S5a and S5b** represent the THz signal and corresponding THz spectrum recorded for the linearly polarized (LP) 800 nm pump beam excitation. We observe that the THz signal of $Bi_2Se_3$ is much smaller than that of $Bi_2Te_3$. The bandwidth of the THz spectrum for both samples is < 3 THz. And the nature of the spectrum for both these TIs is almost similar.

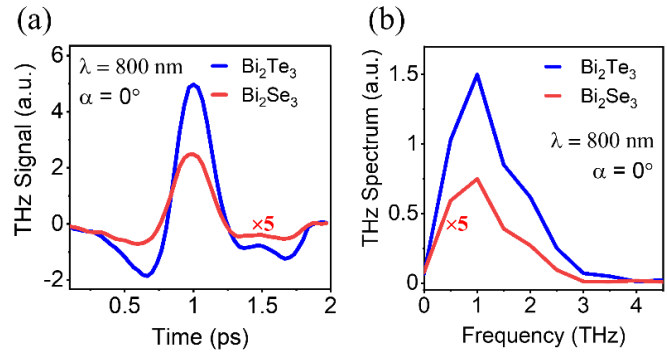

**Figure S5.** a) THz waveform generation from Bi2Te3 and Bi2Se3 single crystalline samples for linearly polarised 800 nm excitation light. b) The Fourier transform of the corresponding THz signal.

## S7. THz electric field calculation

Measurement of the THz signal emitted from $Bi_2Te_3$ and $Bi_2Se_3$ is based on the standard electro-optic sampling scheme as shown in **Fig. 1a of the main text**. The magnitude of the THz electric field based on this scheme can be constructed using the following standard technique[21, 22].

$$\frac{\Delta I}{I_{probe}} = \frac{\omega n^3 r_{41} L}{2c} \cdot E_{THz}^{ZnTe} \cdot (\cos\beta \sin 2\gamma + 2\sin\beta \cos 2\gamma)$$

$$E_{THz}^{ZnTe} = \frac{2c}{\omega n^3 r_{41} L} \cdot \frac{\Delta I}{I_{Probe}} \cdot \frac{1}{(\cos\beta \sin 2\gamma + 2\sin\beta \cos 2\gamma)} \qquad (S7.1)$$

Here, ΔI is the intensity difference of the gating beam measured on the balanced photodiode, $I_{probe}$ is the absolute intensity of the probe beam used in the experiment for electro-optic detection, c is the velocity of light in vacuum, ω is the angular frequency of the probe beam with λ = 800nm. The refractive index and the nonlinear coefficient of the 0.5 mm thick ZnTe nonlinear crystal used in our experimental setup has the standard values of n = 2.85 and $r_{41}$ = 4 pm/V, respectively[23]. Here, β and γ are the angles of the THz beam polarization and probe beam polarization with respect to the (001)-axis of the ZnTe crystal; in the laboratory frame, we call it Y-axis. In our experimental setup we kept β = 90° and γ = 180°.

We calculate the magnitude of the THz electric fields (in mV/cm) corresponding to all the experimentally measured electro-optic signal data values presented in Fig. 3b and 3c under scheme-I and Fig. 3e and 3f under scheme-II using equation 1. We find that calculation gives unrealistically very high estimated values of $E_{pp}^o$ (mV/cm) at higher wavelengths for both samples when $E_{pp}^o$ is estimated under scheme-I. For examples from Fig. 3c under scheme-I, $E_{pp}^o$ (mV/cm) for $Bi_2Se_3$ at 900 nm linearly polarized excitation is 1732700 mV/cm which is almost three orders of magnitude higher than the estimated $E_{pp}^o$ (mV/cm) at 400 nm linearly polarized excitation.

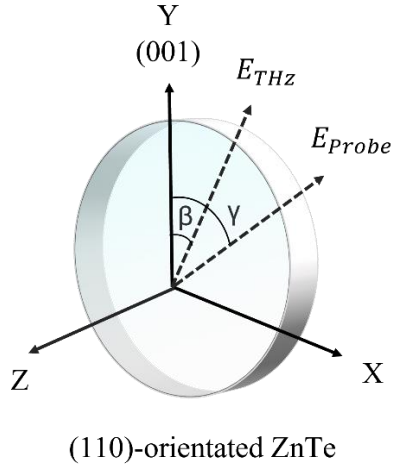

(110)-orientated ZnTe

**Figure S6.** Relative orientation of polarization of pump beam and THz electric field with respect to (001)-axis of the (110-oriented) ZnTe crystal.

**Refereences**